%% file: ChruscielWutteEnergyArxiv.tex
\documentclass[a4paper,11pt]{article}
\usepackage[T1]{fontenc}
\usepackage[utf8]{inputenc}
\usepackage{graphicx}
\usepackage{tensor}
\usepackage{amsmath}
\usepackage{mathtools}
\usepackage{amsfonts}
\usepackage[dvipsnames]{xcolor}
\usepackage{ntheorem}
\usepackage{a4}
\usepackage{hyperref}
\usepackage{cite}
\usepackage{wasysym} 
\usepackage{braket}
\input{macros}

\input{ptcmacros}

\input{macros2}

\input{macrosForEnergy}

\newcommand{\stone}{\blue{\mu}}
\newcommand{\sttwo}{\blue{\nu}}
\newcommand{\stthree}{\blue{\rho}}

\newcommand{\hstone}{\blue{\hat{{\mu}}}}
\newcommand{\hsttwo}{\blue{\hat{{\nu}}}}
\newcommand{\hstthree}{\blue{\hat{{\rho}}}}

\usepackage[normalem]{ulem}

\usepackage{mleftright} \mleftright 

\usepackage{cancel}

 \renewcommand{\red}[1]{{#1}}
 \renewcommand{\blue}[1]{{#1}}

\renewcommand{\rwcheck}[1]{}%
\renewcommand{\ptcheck}[1]{}%
\renewcommand{\checking}[1]{}%

\renewcommand{\bnabla}{\red{\overline{\nabla}}}

\author{Piotr T. Chru\'{s}ciel\thanks{Beijing Institute of Mathematical Sciences and Applications, Huairou, China, and
Center for Theoretical Physics of the Polish Academy of Sciences, Warsaw, Poland} \thanks{
{\sc Email} \protect\url{pchrusciel@cft.edu.pl}, {\sc URL} \protect\url{homepage.univie.ac.at/piotr.chrusciel}}
\\
{Raphaela Wutte}\thanks{Mathematical Sciences and STAG Research Centre, University of
Southampton, Highfield, SO17 1BJ Southampton, United Kingdom}
\thanks{
{\sc Email} \protect\url{rwutte@hep.itp.tuwien.ac.at} }
}

\begin{document}

\title{Holographic is Hamiltonian, relatively}

\maketitle

\begin{abstract}
We show that a relative holographic energy coincides with the relative Hamiltonian energy.
\end{abstract}
 
 \tableofcontents

%
\input{IntroEnergy}

  \input{dim4v3}

\input{holographic}

  \section{Hamiltonian charges}
 
\input{Definition}

\input{BK}
  \input{hamiltonian}

\input{Examples}

\input{Equality}

\input{Comments}

\bigskip
\noindent{\sc Acknowledgements:}
We are grateful to  
Robin Graham, Jaros\l aw Kopi\'nski, and Kostas Skenderis for bibliographical advice. 
RW acknowledges
support from the STFC consolidated grant ST/X000583/1 “New Frontiers in Particle Physics,
Cosmology and Gravity”. 
\\ \\
\noindent{\sc Data availability statement:} No data were created or analyzed in this manuscript.

\appendix 
 
\checking{notes on Wiseman commented out, and on conf transf of spinors, now included in ChruscielWutteSpinors supplementary information } 

\bibliographystyle{amsplain}

\bibliography{ChruscielWutteLetter-minimal,ChruscielWutteEnergy-minimal}\end{document}

%% file: ptcmacros.tex
\newcommand{\hyp}{\blue{\mathcal{S}}}

\newcommand{\beaa}{\begin{eqnarray}}
\newcommand{\eeal}[1]{\end{eqnarray}\label{#1}}

\newcommand{\checking}[1]{\ptc{checking: #1}}

\newcommand{\ourQ}{\red{q}}
\newcommand{\theirQ}{\red{Q}}

\newcommand{\zht}{\red{\tau}}

\newcommand{\theirT}{\red{t}}

\newcommand{\secscri}{\blue{\mathcal{C}}}

\newcommand{\xa}{\red{y^a}}
\newcommand{\xb}{\red{y^b}}

\newcommand{\smalla}{\red{a}}
\newcommand{\smallb}{\red{b}}

\newcommand{\hsmalla}{\red{\hat a}}
\newcommand{\hsmallb}{\red{\hat b}}

%% file: macros2.tex
\newcommand{\hA}{\blue{\hat{A}}}

\newcommand{\zR}{\red{\mathring{R}}}

\newcommand{\myh}{\blue{\gamma}}
\newcommand{\bg}{\backg}
\newcommand{\mnsigma}{\red{s}}
\newcommand{\mnnsigma}{\red{s}}

\newcommand{\scri}{\blue{\mycal I}}
\newcommand{\Scri}{\scri}

\newcommand{\zX}{\red{\mathring{X}}}

\newcommand{\zD}{\mathring{D}}%



%% file: macrosForEnergy.tex
\newcommand{\redOof}{\blue{O(}}

\newcommand{\CHringh}{\blue{\zgamma}}
 \newcommand{\const}{\blue{\text{const}}}
 
 \newcommand{\mham}{\blue{H[\hyp,\partial_t,\backg](g)}}

\newcommand{\dangerous}{}
\newcommand{\mdanger}{\mbox{\red{\dangerous}}}

\newcommand{\eCIBh}{f {\mdanger}}

\newcommand{\zero}{\blue{\hat 0}}
\newcommand{\one}{\blue{\hat n}}
\newcommand{\onet}{\blue{\hat 3}}
\newcommand{\two}{\blue{\hat 1}}
\newcommand{\three}{\blue{\hat 2}}

\newcommand{\chitwo}{\blue{\varphi}}

\newcommand{\bean}{\begin{eqnarray}}

%% file: IntroEnergy.tex
\section{Introduction}

Consider a 
stationary
$(n+1)$-dimensional spacetime $(\mcM,\backg)$, with $\backg$ solving the vacuum Einstein equations with a negative cosmological constant.  (Large families of such metrics exist by~\cite{ACD2,CD4,ChDelayKlingerBH,KlingerBirkhoff}.)
Suppose that $(\mcM,\backg)$ admits a conformal boundary at infinity $\scri$ \emph{\`a la Penrose} with a, say smooth, conformal class of metrics induced on $\scri$. 
Each such metric $\backg$ comes with a family of vacuum, dynamical, metrics $(\mcM, g)$ which share their conformal class at infinity with $\backg$~\cite{Friedrich:aDS,KaminskiFG}. 
For each such metric $g$ one can define both its holographic energy~\cite{Papadimitriou:2005ii} (see also \cite{Henningson:1998gx,deHaro:2000vlm,Balasubramanian:1999re,Skenderis:2000in,Cheng:2005wk})
  and its Hamiltonian energy relative to $\backg$~\cite{ChAIHP}. 
  The aim of this note is to show that a  holographic energy of $g$ 
  defined relatively to $\backg $ coincides with the Hamiltonian energy. 
 This fact  has previously been pointed out in spacetime dimensions $3+1$  in~\cite{ChruscielSimon},
 but  the result here holds in all dimensions $n+1\ge 4$ and for any conformal metric on $\scri$. 
(The case $n+1=3$ requires separate treatment, cf.\ e.g.~\cite{ChWutte} and references therein.)

In this article we are concerned with the Hamiltonian energy as defined in \cite{ChAIHP}, based on the formalism developed in~\cite{KijowskiTulczyjew}. The approach is related to, but does not coincide with, a subset of  the much more extensive treatment in  \cite{Papadimitriou:2005ii} where the holographic charges are considered  from the perspective of the covariant phase space analysis in the spirit of \cite{Lee:1990nz,Wald:1999wa}.

In retrospect, our calculations here give a tacit way of implementing the counterterm-subtraction of \cite{Papadimitriou:2005ii}.

As such, we have assumed a stationary background, as the  Hamiltonian charge associated with a vector field $X$ which is timelike at large distances and is a Killing vector field for the metric  $\backg$ usually has the interpretation of energy. One can similarly define relative Hamiltonian charges for any $\backg$-Killing vector $X$, regardless of its causal character, concluding again that for vacuum metrics $g$ and $\backg$ which share the same conformal structure at $\scri$ the Hamiltonian charge of $g$ relative to $\backg$ equals the difference of holographic charges of $g$ and $\backg$;
see Section~\ref{s2IV26.1} for more comments on this.

%% file: dim4v3.tex
\section{Holographic charges} 

When the matter fields decay sufficiently fast and the spacetime admits a smooth conformal completion at infinity,
there exist coordinates in which the metric can be written in the form  
\begin{align}\label{6325ac}
    g &= x^{-2}
    \big(
     dx^2 + (\zgamma_{AB} + x^2 \gammaextwo_{AB}
     + \ldots
     + x^n \log x \gammalog_{AB} + x^n \gammexn_{AB}
     + \ldots
     ) dy^A dy^B
      \big) \\
      &=: x^{-2} (dx^2 + \gamma_{AB} dy^A dy^B)
    \,,\label{6325bc}
  \end{align}
  where for $n \geq 3$ \cite{deHaro:2000vlm}
  \begin{equation}
    \gammaextwo_{AB} = - \frac{1}{n-2} ({\mathring R}_{AB} - \frac{1}{2(n-1)} {\mathring R} \zgamma_{AB})\,.
  \end{equation}

Recall some results reviewed or derived in~\cite{GrahamAmbient} for Einstein metrics $g$. 
Keeping in mind that $n$ is the dimension of the  boundary,  if $n$ is odd  then 
$$
\gammexn:= \gammexn_{AB} dx^A dx^B
$$
is $\zgamma$-trace-free and $\zgamma$-divergence-free.  If $ n$ is even, then the trace and divergence of  $ \gammexn$ 
are determined locally in terms of the boundary metric, and are nonzero in general.  This is contained in~\cite[Theorem 4.8]{GrahamAmbient}.  The divergence is identified explicitly when $n=4$, and~\cite[Equation~(3.19)]{GrahamAmbient}  identifies the trace explicitly for $n=4,6$.  
For even boundary dimension \(n\ge6\), the full coefficient \(\gammexn\) is trace-free and divergence-free if the chosen boundary metric is locally conformally flat or Einstein. In dimension four, under those same hypotheses, the tensor with both properties is \(\gammaexfour_{AB} -\tfrac14 \gammaextwo_{AC} \zgamma^{CD} \gammaextwo_{DB}\).

Set  
\begin{equation}\label{28I26.1} 
 \zht\equiv 
 \zht_{AB}dx^Adx^B := 
  \big(
\gammexn_{AB}- \frac1n \zgamma^{CD}
\gammexn_{CD} \zgamma_{AB}\big)
  \big|_{x=0} dx^A dx^B
   \,.
\end{equation}
%

Given a conformal Killing vector $X$ of the conformal structure of $\scri$ and a section $\secscri$ of $\scri:=\{x=0\}$, 
when the divergence of $\zht$ vanishes, 
it is natural to define
\emph{holographic-type charges}  as 
 \checking{changed $Q_H$ to $\ourQ$ macro ourQ, and name, check for consistency} 
  \begin{equation}
    \label{holcharge}
    \ourQ[\secscri,X](g) = - \int_{\secscri}
     \zgamma^{AC}\zht_{CB}X^B dS_A 
    \,,
    \end{equation}
where $dS_A$ is given by $\sqrt{-\det \zgamma }\, \partial_A\, \rfloor dx^1\wedge \cdots \wedge dx^n$.
Then $\ourQ$ is the same for all sections of $\scri$. 
We emphasise that our calculations below show  that $\ourQ$ arises naturally from a Hamiltonian analysis of the theory regardless   of the vanishing of the divergence of $\zht$, for any vector field $X$ 
for which the Hamiltonian volume integrals converge; compare Section~\ref{s2IV26.1}.

%% file: holographic.tex
Recall that the \emph{holographic energy momentum tensor $\theirT$} is defined as \cite[Equation~(17)]{Skenderis:2000in}
\checking{changed to $\theirT$, macro theirT for consistency with the letter, watch out for consistency}
  \begin{equation}
  \theirT_{AB} = \frac{n }{16 \pi G} (\gammexn_{AB} + X^{(n)}_{AB})
  \,,
  \end{equation}
 with $X^{(n)}_{AB} = 0$ for odd $n$. 
 The field $X^{(n)}_{AB}$ depends only upon the fields  $\zgamma$ and $ \gammexa$ with $1\le a < n$ and their derivatives, 
  it vanishes in odd space dimensions, and
  its exact form depends on the spacetime dimension  as explained in the paragraph following Equation~(1.3) of \cite{deHaro:2000vlm} 
 \footnote{In principle, $h$ enters in the  formula for the energy-momentum as well, but the authors have a holographic argument to remove it, see 
  \cite[text under (3.13)]{deHaro:2000vlm}.}. 
Hence, upon requiring that the matter fields decay to sufficiently high order at $\scri$,   it is sufficient to know $\zgamma$ and $\gammexn$ to calculate $\theirT_{AB}$.

  The charges $\theirQ $ are then defined as \cite[Equation~(3.17)]{Cheng:2005wk} (compare~\cite{deHaro:2000vlm,Compere:2008us})
  \begin{equation}
\theirQ [\secscri,X](g) = - \int_{\secscri}
     \zgamma^{AC}\theirT_{CB}X^B dS_A 
\,.
\label{skencharge}
  \end{equation}
  Hence $\theirQ $ differs from $\ourQ$ by a multiplicative factor and, for even $n$, a shift depending only upon the metric $\zgamma$ and its derivatives.  
  According to~\cite[Equation~(3.16)]{Cheng:2005wk} the tensor field $\theirT_{AB}$ is divergence free in any dimension.

  \checking{ if you know some other refs, add them here}

  Alternatively, the energy-momentum tensor can also be expressed as 
   \cite[Equation (49)]{Papadimitriou:2004ap} (see also\cite[Equation (3.15)]{Cheng:2005wk})
  \begin{equation}
  \theirT_{AB} = - \frac{1}{8 \pi G} \left( K_{(n) AB} - (\zgamma^{CD} K_{(n) CD}) \zgamma_{AB}\right)\,,
  \end{equation}
  where $K_{(n)}$ are certain expansion coefficients of the extrinsic curvature of constant $x$ slices in terms of eigenfunctions of the so-called dilation operator.

While the divergence of the tensor $\tau$ defined in \eqref{28I26.1} does not vanish for general conformal boundaries at infinity and even $n$,
the tensor $\tau$  is convenient in the analysis below due to its simple form. The corresponding charge \eqref{holcharge} differs from \eqref{skencharge} (aside from a multiplicative factor) only by a term which cancels in the relative charge $Q[\secscri,X](g) - Q[\secscri,X](\bar g)$ between two metrics $g$ and $\bar{g}$ sharing the same conformal infinity.

%% file: Definition.tex
 
Given a vector field $X$ and a spacelike hypersurface $\hyp\subset \mcM$ with boundary $\partial\hyp$ included in $\scri$, consider the Hamiltonian charges as derived in~\cite{ChAIHP}   
\begin{equation}
H[\hyp,X,\backg](g) =
 \frac 12  \int_{\partial {\hyp}} E^{\stone \sttwo} \eta_{\stone \sttwo} \,,
 \label{4II26.1}
\end{equation}
with $\stone,\sttwo\in\{0,\ldots,n\}$, where $\backg$ is a background metric with respect to which the mass will be defined.
Further $\eta_{\stone \sttwo} = \partial_\stone \rfloor \partial_\sttwo \rfloor 
(d t \wedge dx^1 \wedge ... \wedge dx^{n-1} \wedge dx)$
and 
\begin{equation}
E^{\stone \sttwo} = E^{\stone \sttwo}_K - E^{\stone \sttwo}_K \vert_{g = \backg}+ \frac{1}{8\pi} X^{[\stone} Z^{\sttwo]} \sqrt{|\det \backg|}
 \,,
\end{equation}
with 
\begin{equation}
   E^{\stone \sttwo}_K = \frac{\sqrt{|\det g|}}{16 \pi}(\nabla^\stone X^\sttwo - \nabla^\sttwo X^\stone)
   \,,
\end{equation}
\begin{equation}
Z^\stone = e(g^{\stone \sttwo} C^\stthree_{\sttwo \stthree} - g^{\sttwo \stthree} C_{\sttwo \stthree}^\stone) 
 \,,
 \qquad C^\stone_{\sttwo \stthree} = \Gamma^\stone_{\sttwo \stthree} - {\bar \Gamma}^\stone_{\sttwo \stthree}
 \,,
\end{equation}
and 
$e = \sqrt{\det (\backg^{\stone \stthree} g_{\stthree \sttwo})}$, where $\bar \Gamma^\stone_{\sttwo \stthree}$ denotes Christoffel symbols of the metric $\bar g$.  

%% file: BK.tex
\subsection{Asymptotically Birmingham-Kottler metrics}
 \label{ss6II26.2}

In this section we consider $(n+1)$-dimensional Lorentzian metrics $g$  which  asymptote to the background metric $\backg$ given by  
\begin{equation}\label{22IV18.11}
  \backg= \red{-} \left( \beta +\frac{r^2}{\ell^2}\right) dt^2 + \frac {dr^2} {\left( \beta +\frac{r^2}{\ell^2}\right)} + r^2  \CHringh
 \,,
 \quad
 \CHringh= \CHringh _{\smalla \smallb}(x^C) d\xa d\xb
 \,,
\ee
with $\beta \in\{0,\pm1\}$, where $(x^\stone)\equiv(t,y^a,r)$,
and where $\ell^2$ is a constant inversely proportional   to the cosmological constant.

The $\backg$-Killing vector $X$ is taken to be  $\partial_t$, and the spacelike hypersurface is  $\hyp=\{t=\const\}$.

We use the following $\backg$--orthonormal frame:
\begin{equation}
  \label{22IV18.41}
  \eCIBh_{\zero}= \frac{1}{\sqrt{\beta + \frac{r^2}{\ell^2}}}\partial_t\,, \quad
  \eCIBh_{\one}=  {\sqrt{\beta + \frac{r^2}{\ell^2}}} \partial_r\,, \quad
  \eCIBh_{\hsmalla } = \frac 1r \chitwo _{\hsmalla }\,,
\end{equation}
where $\chitwo _{\hsmalla }$ is an orthonormal (ON) frame for the metric $\CHringh$.  
Here, and in what follows in the current section, we use 
$$
 \smalla \in \{1,...,n-1\}
 \,,
$$
and we shall use hatted indices to denote the components of a tensor field
in the frame $\eCIBh_{{\hstone }}$ defined in \eqref{22IV18.41}.

Let the tensor field $\ehere^{\stone \sttwo}$ be defined by the formula 
\checking{ note that emunu is NOT the same as  hmunu}
\begin{equation}
  \label{mas3_new}
  \ehere^{\stone \sttwo}
   := g^{\stone \sttwo}-\backg^{\stone \sttwo}
	\,,
\end{equation}
thus
$$
 \ehere^{\stone \sttwo}\partial_\stone \otimes \partial_\sttwo = \ehere^{{\hstone }{\hsttwo }}\eCIBh_{{\hstone }}\otimes
    \eCIBh_{{\hsttwo }}
 \,.
$$

 We will say that a metric is \emph{asymptotically Birmingham-Kottler} if  there exists $\epsilon>0$ such that in the frame \eqref{22IV18.41} it holds  
\begin{equation}
  \label{hamfalof_new1}
  \ehere^{\hstone \hsttwo }= \redOof r^{-n/2-\epsilon})\,, \quad 
   \eCIBh_{\hstthree}(\ehere^{\hstone \hsttwo })= \redOof r^{-n/2-\epsilon})
  \,,
   \quad
   \det (g^{\hstone \hsttwo }  )   = -1+\redOof r^{-n-\epsilon})\,.
\end{equation}
For matter fields with finite integrated energy, this guarantees convergence of the Hamiltonian mass and is optimal when pointwise-decay conditions are imposed.

\checking{you decide whether you want to include BKcalculations here} 
A calculation leads to the  following simple expression for the \emph{Hamiltonian mass $\mham$} of asymptotically Birmingham-Kottler metrics:
\begin{eqnarray}
	\label{MHamBh}
	\mham
	&=&
	\lim_{R\to\infty} \frac {R^n} {16\pi}
	\int_{\hyp \cap\{r=R\}}
		\left[
		\left(\frac{1}{\ell^2}+  \frac{\beta}{r^2}  \right)
    \left({\frac {r \partial \big(\backg_{{\hsmalla }{\hsmallb }}\ehere^{{\hsmalla }{\hsmallb }} \big)}{\partial r}} - (n-1)
   \ehere^{\one\one}  \right)
   \right.
		\nonumber \\
		&&
	\phantom{xxxxxxxxxxxxxxxxxxxxx}
\left.
		+
		 \frac{\beta}{ {r^2} }  \backg_{{\hsmalla }{\hsmallb }}\ehere^{{\hsmalla }{\hsmallb }}
  \right] d^{n-1}\mu_{{\CHringh}}
	\,.
\end{eqnarray}
In spacetime dimension $n+1=4$ this simplifies to the expression given in~\cite{ChruscielSimon}:
\begin{eqnarray}
	\label{MHamBh3}
	\mham
	&=&
	\lim_{R\to\infty} \frac {R^3} {16\pi\ell^2}
	\int_{\hyp \cap\{r=R\}}
		\left[
		r \frac{ \partial \ehere^{{\two}{\two}} }{\partial r}
		+
		r \frac{ \partial \ehere^{{\three}{\three}} }{\partial r}
		-
		2 \ehere^{\onet\onet}
  \right] d^{n-1}\mu_{{\CHringh}}
	\,.
	\phantom{xxx}
\end{eqnarray}
If in addition to \eqref{hamfalof_new1} we assume that
\begin{equation}
  \label{hamfalof_new1strong}
  \ehere^{\hstone \hsttwo}= \redOof r^{-n} )\,, \quad \eCIBh_{\hstthree }(\ehere^{\hstone \hsttwo})= \redOof r^{-n})
  \,,
\end{equation}
which is  the fall-off rate for Birmingham metrics, Equation~\eqref{MHamBh} can be rewritten in a form similar to \eqref{MHamBh3} in higher dimensions as well:
\begin{eqnarray}
	\label{MHamBhstrong}
	\mham
	&=&
	\lim_{R\to\infty} \frac {R^n} {16\pi\ell^2}
	\int_{\hyp \cap\{r=R\}}
		\left[
   {r   \backg_{{\hsmalla }{\hsmallb }}\frac {\partial \ehere^{{\hsmalla }{\hsmallb }} }{\partial r}} - (n-1)
   \ehere^{\one\one}  
  \right] d^{n-1}\mu_{{\CHringh}}
	\,.
   \phantom{xxxxxx}
\end{eqnarray}

%% file: hamiltonian.tex
\subsection{General smooth $\Scri$}
 \label{ss3IV26.1}
\checking{beginning of hamiltonian.tex}

Consider $(n+1)$-dimensional metrics of the form 
\begin{equation}
g_{\stone \sttwo}(x,x^A) = \backg_{\stone \sttwo}(x,x^A) +x^{n-2} h_{\stone \sttwo}(x,x^A) 
  \,,
   \label{10IX25.1}
\end{equation}
with $\stone, \sttwo \in\{0,\ldots,n\}$, where $\backg$ is a background metric with respect to which the mass will be defined. 
We use a Fefferman-Graham coordinate system 
$$(x^\stone) 
\equiv ( y^A,x)\equiv (t,x^1,\ldots,x^{n-1},x) 
    \equiv (x^0,x^1,\ldots,x^{n-1},x) 
 \,,
$$
thus
$$
\mbox{$g_{xx} = \backg_{xx} = x^{-2}$,\quad $g_{xA} = \backg_{xA} = 0$, }
$$
and
\begin{equation}
 \label{14II26.1}
g^{\stone \sttwo} = \backg^{\stone \sttwo} - x^{n-2} h^{\stone \sttwo} + ...  
\,,
\end{equation}
with the indices on $h_{\stone \sttwo}$  raised and lowered with $\backg_{\stone \sttwo}$. 
(Note that in the asymptotically BK case the leading order terms of the field $e^{\mu\nu}$ of  \eqref{mas3_new} coincide with the leading order terms of the field $-x^{n-2}h^{\mu\nu}$.)

Let us assume that both metrics $g$ and $\backg$ are vacuum.  
Since $g$ and $\backg$ share the same smooth conformal infinity, their asymptotic expansions at $\{x=0\}$ to the order made explicit in \eqref{10IX25.1} will  coincide, including   logarithmic terms if any. 
It follows that the coordinate components  
$h_{\stone \sttwo}$ of the tensor field $h_{\stone \sttwo}dx^\stone dx^\sttwo $ defined in \eqref{10IX25.1}  satisfy  $h_{x \stone}\equiv 0$ and 
\begin{equation}\label{26VIII25.3}
  h_{AB} =  
 \mathring h_{AB} + O(x\ln x)
\,,
 \ \mathring h_{\stone \sttwo}:= h_{\stone \sttwo}|_{x=0} 
  \,,
  \ 
  \partial_x   h_{AB} = O(\ln x)
  \,,
  \ 
  \partial_A h_{BC} = O(1)
  \,. 
\end{equation}

We note that much weaker asymptotic conditions would suffice for finite, well defined Hamiltonian charges. However, the formulae below would need a careful reexamination.
This is irrelevant in the current context in any case, as we want to determine the Hamiltonian charges for metrics which have a well defined holographic mass.

We also note that matter fields which decay sufficiently fast will not change the calculations here.
 
Denoting by $\mathring D$  the covariant derivative operator of the metric $\mathring \gamma_{AB}$ of \eqref{6325ac}, one finds   
\begin{align}
g^{tA} C^\stthree_{A \stthree} 
 &= 
  \frac{x^{n-2}}{2} g^{tA} \backg^{BC}\bnabla_A h_{BC} 
    +O(x^{2n+2}) 
 \nonumber 
\\
 & = 
 \frac{x^{n+2}}{2} \zgamma^{tA} \zgamma^{BC}\bnabla_A h_{BC} 
 \big(1+ O(x^2)
 \big)
+O(x^{2n+2})  
 \nonumber
\\
 & = 
 \frac{x^{n+2}}{2} \zgamma^{tA} \zgamma^{BC}\zD_A \zhhere_{BC}  
+O(x^{n+3}) 
= O(x^{n+2})
\,, 
\\
    g^{\sttwo \stthree} C^t_{\sttwo \stthree} &=  
  x^{n +2}
  \gamma^{AB}\backgamma^{tC} 
  \left(
   \bnabla_A h_{BC} - \frac{1}{2} \bnabla_C h_{AB} 
     \right) 
       +O(x^{2n+2})  
        \nonumber
\\
    &=  
  x^{n +2}
  \zgamma^{AB}\zgamma^{tC} 
  \left(
   \zD_A \zhhere_{BC} - \frac{1}{2} \zD_C \zhhere_{AB} 
     \right)
     \big(1+ O(x)\big)
       +O(x^{2n+2})  
        \nonumber 
\\
    &=  
  x^{n +2}
  \zgamma^{AB}\zgamma^{tC} 
  \left(
   \zD_A \zhhere_{BC} - \frac{1}{2} \zD_C \zhhere_{AB} 
     \right) 
       +O(x^{n+3}) 
= O(x^{n+2}) 
\,.
\end{align}
Since $e = 1 + O(x^{\red{n}})$, we obtain
\begin{equation}
Z^t = 
  x^{n +2}
  \zgamma^{AB}\zgamma^{tC} 
  \left(
    \zD_C \zhhere_{AB} 
    -
    \zD_A \zhhere_{BC} 
     \right) 
       +O(x^{n+3}) 
= O(x^{n+2}) 
 \,.
\end{equation}
Next,
\begin{align}
g^{xx} C^\stthree_{x \stthree} &= \frac{ {n}}{2}  x^{n-1} \backg^{AB} h_{AB}
 + O(x^{n+2})
  = \frac{n}{2} x^{n+1} \zgamma^{AB} \zhhere_{AB} + O(x^{n+2})
  \,,
    \label{eq177251}
\\
    g^{\sttwo \stthree} C^x_{\sttwo \stthree} &= - \frac{(n-2)}{2} x^{n-1} \backg^{AB} h_{AB}
+ O(x^{\red{n+2}})
 \nonumber
 \\
 &= 
- \frac{(n-2)}{2} x^{n+1} \zgamma^{AB} \zhhere_{AB}
+ O(x^{\red{n+2}})
\,,
\end{align}
so that
 \ptcheck{9IX, RW: crosschecked Zx from scratch}
\begin{equation}
Z^x = \red{(n-1)} x^{n+1} \zgamma^{AB} \zhhere_{AB}
 + O(x^{\red{n+2}})
\,.
\end{equation}
Furthermore, 
\ptcheck{5IX; \\ -- \\ rw: checked}
\begin{align}
E^{xt}_K  &= 
 \frac{\sqrt{|\det g|}}{16 \pi}(\nabla^x X^t - \nabla^t X^x) 
 \nonumber
\\ &= 
 \frac{\sqrt{|\det g|}}{16 \pi}g^{x \stone} g^{t \sttwo} (\partial_\stone X_\sttwo  -   \partial_\sttwo  X_\stone) 
 \nonumber
\\ &= 
 \frac{\sqrt{|\det g|}}{16 \pi}g^{x x} g^{t \sttwo} (\partial_x X_\sttwo  -   \partial_\sttwo  X_x) 
 \nonumber
\\
    &=  \frac{\sqrt{|\det g|}}{16 \pi} g^{tA} \left( 
   x^2 \partial_x(g_{AB} X^B) - \partial_A X^x
    \right)\,.
    \label{5IX25.1}
\end{align}

 To continue, we assume that $X$ is a Killing vector of $\backg$, possibly  up to error terms which decay fast:
\begin{equation}\label{26VIII25.5}
  \bnabla_\stone X_\sttwo + \bnabla_\sttwo X_\stone = o(x^{\mnnsigma-2})
  \,, 
\end{equation}
where $o(x^{\mnnsigma-2})$ is meant in   
terms of tensor components in Fefferman-Graham coordinates. 
The reader should think of the exponent $\mnsigma$ as the one needed for finiteness of the integral in \eqref{4II26.1}, with possibly $o$ replaced by $O$, but in the current calculations we do not make any assumptions about $\mnsigma$ until explicitly indicated otherwise.

For conformally smooth backgrounds $\bar g$, a $\bar g$-Killing   vector field  $X$ necessarily extends smoothly to the conformal boundary at infinity $\partial \mcM$, with the extension tangent to the conformal boundary, its restriction to the boundary being a conformal Killing vector of the conformal metric on $\partial \mcM$.  We will therefore assume that $X$ extends smoothly and tangentially to a conformal Killing vector of the conformal boundary. 
One can without loss of generality assume that the coordinates $(x,y^A)$ are simultaneously Fefferman-Graham coordinates both for the background $\bg$ and the metric of interest $g$.
Writing  
\begin{equation}
  \bg  = x^{-2}
  (dx^2 + \backgamma_{AB} dy^A dy^B)
  \label{eqgan}\,,
\end{equation}
where 
$$
  \backgamma_{AB}(x,y^C)
  = \zgamma_{AB}(y^C) + O(x^2)
  \,,
$$  
the asymptotic Killing equations in Fefferman-Graham coordinates
\begin{equation}
 X^\stone \partial_\stone \backg_{\sttwo \stthree} + \partial_\sttwo X^\stone \backg_{\stone \stthree}
  + \partial_\stthree X^\stone \backg_{\stone \sttwo} = 
  o(x^{\mnsigma-2})
  \label{26VIII25.9}
\end{equation}
with $\sttwo \stthree=xx$ read
 \ptcheck{12IV26; no upper bound on $\mnsigma$}
\begin{align}\label{26VIII25.7}
  -2 x^{-3} X^x  
  = \  & -  2 x^{-2}\partial_x X^x + 
  o(x^{\mnsigma-2})
   \,.
\end{align}
Integrating in $x$ we conclude that, for  $\mnsigma \ne 0$ (the case $\mnsigma=0$ would need separate and irrelevant attention, as it would introduce annoying log terms),
\checking{2IX; it would be logical to change the notation for mathring X to something like X overscript [1], but ugly; though it is consistent with behaviour in norm \\ -- \\ rw: overscript [1] seems a bit heavy}
\begin{align}\label{26VIII25.8}
    X^x  
  = \  & x \zX^x + 
  o(x^{\mnsigma+1})
   \,, 
   \ \partial_x \zX^x=0
   \,.
\end{align}
Inserting this into \eqref{26VIII25.9} with 
 $\sttwo \stthree=xA$  gives  
\ptcheck{2IX25? and 12 IV 26} 
\begin{equation}\label{16VIII25.9}
  X^A = \zX^A -  \blue{\frac{x^2}{2} } \mathring \myh^{AC} \partial_C \zX^x 
   \big(1 + O(x^{2})
  \big) + o(x^{\mnsigma+1})
   \,,
   \ \partial_x \zX^A=0
   \,,
    \ 
   \mnsigma \not\in \{-1,0\}
   \,,
\end{equation}
where the new restriction on $\mnsigma$ arises again to avoid dealing with 
further terms involving $\log x$.  
 
Returning to \eqref{5IX25.1},  
for  $s>0$  
we can write
\ptcheck{rw: checked}
\begin{align}
E^{xt}_K |_{g=\bg}
    = \ &  
     \frac{\sqrt{|\det \bg|}}{16 \pi} \bg^{tA} \left( 
   x^2 \partial_x(\bg_{AB} X^B) - \partial_A X^x
    \right)
    \,,
     \nonumber
\\     
E^{xt}_K 
    = \ &  \frac{\sqrt{|\det g|}}{16 \pi} g^{tA} \left( 
   x^2 \partial_x(g_{AB} X^B) - \partial_A X^x
    \right)
     \nonumber
\\      
    = \ &  
E^{xt}_K |_{g=\bg}
    +
     \frac{\sqrt{|\det \zgamma|}}{16 \pi} 
     \Big[ \frac {x}2 \zgamma^{CD}\zhhere_{CD}
     \zgamma ^{tA} \left( 
   x^2 \partial_x(x^{-2} \zgamma_{AB} \zX^B) - 
   \red{x} \partial_A \zX^x
    \right) 
     \nonumber
\\      
    &
     -  x\,
     \zgamma^{tC}\zgamma^{AD}\zhhere_{CD} \left( 
   x^2 \partial_x(x^{-2} \zgamma_{AB} \zX^B) -    \red{x} \partial_A \zX^x
    \right) 
     \nonumber
\\      
    & +    x^{-n+1}
     \zgamma ^{tA} \left( 
   x^2 \partial_x(x^{n-2} h_{AB} \zX^B)
    \right)
    \Big]
     + o(1)
     \nonumber 
\\      
    = \ &  
E^{xt}_K |_{g=\bg}
    +
     \frac{\sqrt{|\det \zgamma|}}{16 \pi} 
     \Big[ 
    \big(- \zgamma^{CD}
     \zgamma ^{tA} + \red{2}
     \zgamma^{tC}\zgamma^{AD} 
     \big)
     \zhhere_{CD} \zgamma_{AB} \zX^B 
     \nonumber
\\      
    & +   (n-2) 
     \zgamma ^{tA} \zhhere_{AB} \zX^B 
    \Big]
     + o(1)
 \,.
\label{5IX25.20}
\end{align}
Hence 
\rwcheck{checked}
\begin{align} 
E^{xt}_K -
E^{xt}_K |_{g=\bg}
    = \ &   
     \frac{\sqrt{|\det \zgamma|}}{16 \pi} 
     \Big[ 
    -  \zgamma^{CD} \zhhere_{CD}  \zgamma_{AB}
    +   n
      \zhhere_{AB}  
    \Big]  \zgamma ^{tA} \zX^B 
     + o(1) 
 \,.
\label{5IX25.2}
\end{align}
When
\begin{equation}\label{26VIII25.2}
\mbox{
 $\partial\hyp =\{t=0\}\cap \{x=0\}$ 
and $\mnsigma>0$ 
 }
 \end{equation}
 we obtain  
\begin{align}\label{26VIII25.1}
H = \  &
 - 
   \int_{\partial {\hyp}} E^{xt} \,
 \red{d^{n-1} y} 
 \nonumber
\\
 = \ &
\frac{1}{16 \pi}
 \int_{\partial \hyp}
    \Big( \big[ 
    \zgamma^{CD} \zhhere_{CD}  \zgamma_{AB}
    -   n
      \zhhere_{AB}  
    \big]  \zgamma ^{tA} \zX^B 
    \nonumber
\\ 
  &   
  \phantom{xxxxxxxxxxxxxxxx}
  +
   \red{(n-1)} \zX^t\zgamma^{AB} \zhhere_{AB} 
    \Big)
    \sqrt{|\det \zgamma|} \red{d^{n-1} y} 
 \nonumber
\\
 = \ &
\frac{n}{16 \pi}
 \int_{\partial \hyp}
    \Big( 
      \zgamma^{CD} \zhhere_{CD}  \zgamma_{AB}
    -   
      \zhhere_{AB}  
    \Big)
 ( \zD^{A}t )\zX^B  
    \sqrt{|\det \zgamma|} \red{d^{n-1} y} 
 \,,
\end{align}
where the overall minus sign arises from the fact that the \emph{outer} pointing normal to the level sets of $x$   is $-x\partial_x$.  
\checking{ennd of hamiltonian.tex}

%% file: Examples.tex
\subsection{Examples}
\label{subsec:ex}
As an example, consider the case where $\zX^\stone \partial_\stone = \zX^t \partial_t$, 
with $H$ usually interpreted as the total energy if moreover $\zX^t=1$. 
One obtains 
\rwcheck{checked }
\begin{align}\label{26VIII25.1a}
H   
= \  & 
\frac{1}{16 \pi}
 \int_{\partial \hyp}
    \Big(  
    -   n
      \zhhere_{At}  \zD^A t
  +
    \blue{n} \zgamma^{AB} \zhhere_{AB} 
    \Big)\zX^t
    \sqrt{|\det \zgamma|} \red{d^{n-1} y}     
 \,.
\end{align}
As a consistency check, let us apply \eqref{MHamBhstrong} to metrics of the form \eqref{14II26.1}. For definiteness assume that $\ell=1$, and for $r$ large 
 define $x$  through the equation
\begin{equation}\label{14II26.3}
  \frac{dx}{x} = - \frac{dr}{\sqrt{\beta+r^2 }}
   \,.
\end{equation}
We choose the solution with the asymptotics 
\begin{equation}\label{14II26.4}
  x = \frac{1}{r}- \frac{\beta}{4r^3} +O(r^{-5})
   \,.
\end{equation}
Then $e^{\hstone \hsttwo} = - x^{ n {\red { - 2}}} h^{\hstone \hsttwo} + O(x^{n+2}) $ 
with
$$
\mbox{
$ h^{\one\one} =0$, $\red{{ h}}^{\hat A \hat B} = O(x^2)$,
$r\partial_r =-\big(x +O(x^3)\big)\partial_x$,  
}
$$
and
 \eqref{MHamBhstrong}  becomes, with $(f_{\hat A}) \equiv  (f_{\hat 0},f_{\hsmalla}) $, 
\begin{eqnarray}
	\mham
	&=&
	  \lim_{x\to0} \frac {n} {16\pi }
	\int_{\hyp \cap\{r=R\}}
 \red{x^{-2}}
		      \backg_{{\hsmalla }{\hsmallb }} { {h}}^{{\hsmalla }{\hsmallb }}   d^{n-1}\mu_{{\CHringh}}
\nn
\\
&=&
	  \lim_{x\to0} \frac {n} {16\pi }
	\int_{\hyp \cap\{r=R\}} 
  \red{x^{-2}}
\big(
		      \backg_{\hat A \hat B }  { h}^{\hat A \hat B} 
    +
     { \red{ h}}_{\hat 0 \hat 0}
     \big)
        d^{n-1}\mu_{{\CHringh}}
	\,.
   \phantom{xxxxxx}
 \label{14II26.2}
\end{eqnarray}
The volume condition in \eqref{hamfalof_new1}, namely  
\begin{equation}
  \label{16II26.p1} 
   \det (
    \red{\backg^{\hstone \hsttwo }+}
   e^{\hstone \hsttwo }) {+} 1 = \redOof r^{-n-\epsilon})
   \,,
\end{equation}
requires
\begin{equation}\label{16II26.p2}
\red{x^{-2}} \backg_{\hat A \hat B }  { \red{ h}}^{\hat A \hat B}|_{x=0} \equiv \zgamma^{CD} \zhhere_{CD}  =0 \,.
\end{equation}
When $\backg$ is the Birmingham-Kottler metric \eqref{22IV18.11} with $\ell=1$,  
and if $X=\partial_t$ we recover \eqref{26VIII25.1}. 

As another example, let $g$  
be a perturbation of a  Siklos wave $\backg$,  where   
 \checking{there are two issues here: if $\zf$ vanishes, we have a perturbation of BK, can we have nontrivial such perturbations with positive energy here? and if $\zf$ does not vanish, does one need to adapt the calculations above to include matter fields?}
\begin{equation}
    \label{conformaltopp}
    \backg = x^{-2} \big( -2 du ds + f(s,x,\red{y^a}) ds^2 + dx^2 +(dy^1)^2 + \ldots 
     + (dy^{n-2})^2
        \big) 
    \,,
\end{equation}
with $a\in\{1,\ldots,n-2\}$ and 
\begin{equation}
  \label{fexp}
    f(s,x,\red{y^a}) = \zf (s, y^a) + x^2 \fexextwo(s,x,\red{y^a}), 
\end{equation}
where  both $\zf$ and $\fexextwo $ are smooth on the conformally completed manifold.
The metric $\backg$ satisfies the vacuum Einstein equations with negative cosmological constant $\Lambda = -n(n-1)/2$ 
if and only if the null energy density $\bar \rho$ vanishes,  where
\begin{align}\label{13VI25.r1}
 \bar \rho:= \ & -\frac{1}{2}
 \Big(\big(\partial_{y^1}\big)^2 + \ldots  + \big(\partial_{y^{n-2}}\big)^2
   +  \partial_x^2  - \frac{n-1}{x}  \partial_x
   \Big) f
   \nn
   \\
   =\ &
   \frac{1}{x^2} \left(x \partial_x f - \frac{1}{2} 
\Delta_{\mathbb{H}^{n-1}} f\right)
   \,,
\end{align}
 where we used the Laplacian $\Delta_{\mathbb{H}^{n-1}}$ of the $(n-1)$-dimensional hyperbolic metric
 \begin{equation}
 x^{-2}\big(dx^2 + (dy^1)^2 + \ldots \blue{+} (dy^{n-2})^2
  \big)
 \,. 
 \end{equation}
Indeed, in $n+1$ space-time dimensions we have  (cf.\ \cite[p.22]{Hirsch:2025usn})
  \begin{equation}
   \label{1III26.1}
G\blue{_{ss}} - \frac{n (n-1)}{2} g\blue{_{ss} }=  \bar \rho \,.
 \end{equation}
with the remaining components of the energy-momentum tensor vanishing. 

\checking{some comments on the equation commented out}

As an example, consider the equation $\bar\rho=0$ in $(3+1)$-dimensions with $f(s,x,\red{y}) = \Re (\frak{f}(x) e^{i\red{l} y})$, $\red{l} \in \mathbb{Z}$. We then have 
 \begin{equation}
 \bar \rho \equiv\frac{\red{l}^{2}}{2} \frak{f}(x) + \frac{1}{x}\frac{d \frak{f}(x)}{dx} - \frac{1}{2}\frac{d^{2} \frak{f}(x)}{dx^{2}} = 0 \,.
 \end{equation}
 The solutions to this equation are 
 \begin{equation}
\frak{f}(x)= \frac{c_1 (\sinh (\red{l} x)-\red{l} x \cosh (\red{l} x))}{\red{l}} + \frac{c_2 (\red{l} x \sinh (\red{l} x)-\cosh (\red{l} x))}{\red{l}}
\,.
 \end{equation}
 It follows that general vacuum Siklos waves can be built as finite or infinite sums of such solutions. It would be interesting to analyse whether or not any of the resulting spacetimes have good global properties.
 
\checking{spherical.tex moved to ChruscielWutteSpinors}
 In $(3+1)$-dimensions the boundary metric $\lim_{x\to0} x^2 \backg|_{dx=0}$ has vanishing Ricci scalar, with the only non-vanishing component of its  Ricci tensor $\mathring R_{AB}$ being
$\zR_{ss}=- \frac{1}{2} \partial_y^2 \zf$, and the only non-zero component 
of   its Cotton  tensor $\mathring C_{AB}$ being
 $\mathring C_{ss}= \frac{1}{2}  \partial_y^3 \zf$. 
Hence a three-dimensional boundary metric is  conformally flat if and only if $\partial_y^3 \zf \equiv 0$. 
 \checking{conjecture: if the boundary is not conformally flat there is a matter field contribution to the calculations of the energy, which then need revising ....
}
%

 We have  
 $\det \zgamma  = - 1$, 
 $$
  \zgamma^{tA} = 
   \zD^A t
   =
   \zgamma^{BA}\frac{\partial t}{\partial x^B } =  
   - \frac{\partial t}{\partial s } \delta^A_u
    - 
    \zf 
    \frac{\partial t}{\partial u } 
    \delta^A_u 
   - 
   \frac{\partial t}{\partial u } \delta^A_s
   + 
   \red{\sum_{a=1}^{n-2}}
   \frac{\partial t}{\partial y^{\red{a} }} \delta^A_{y^{\red{a}}}
  \,,
 $$
  and $X = \partial_u$ so that $\zX^t= \partial t / \partial u$. 
%
 For this background \eqref{26VIII25.1} reads   
\begin{equation}
  \label{enden2}
H = \frac{n}{16\pi} \int_{\partial \hyp}
\left(-   \zhhere_{Au}\zD^{A} t + 
 \zgamma^{AB} \zhhere_{AB} \frac{\partial t}{\partial u}
\right)  ds \, d^{n-2} y
 \,,
\end{equation}
where now the coordinates on $\scri$ are $(u,s,y^a)$, and $\partial \hyp$ is defined implicitly by the equation 
$t(u,s,y^a)=0$. When $\mathring f >0$ a possible choice, compatible with the spacelike character of $\hyp$, is $\partial\hyp=\{u=0\}$,
i.e. $t|_\scri = u$.

%% file: Equality.tex
\section{Equality}

To conclude, let 
\begin{align}\label{6325a}
    g &=
     x^{-2}
    \big(
     dx^2 + (\zgamma_{AB} + x^2 \gammaextwo_{AB}
     + \ldots
     + x^n \log x \gammalog_{AB} + x^n \gammexn_{AB}
     + \ldots
     ) dy^A dy^B
      \big) 
\\
    \backg &=
     x^{-2}
    \big(
     dx^2 + (\zgamma_{AB} + x^2 \overline{\gammaextwo}_{AB}
     + \ldots
     + x^n \log x \overline{\gammalog}_{AB} + x^n \overline{\gammexn}_{AB}
     + \ldots
     ) dy^A dy^B
      \big) 
      \nonumber
\\ &=
     x^{-2}
    \big(
     dx^2 + (\zgamma_{AB} + x^2 \gammaextwo_{AB}
     + \ldots
     + x^n \log x \gammalog_{AB} + x^n \overline{\gammexn}_{AB}
     + \ldots
     ) dy^A dy^B
      \big)       
      \,.
  \end{align}
Assuming that the traces of $\gammexn_{AB}$ and $\overline{\gammexn}_{AB}$ coincide
(which, as already pointed out, is the case for the metrics under consideration
 when the matter fields decay sufficiently fast),  we have  
  \begin{equation}\label{28I26.9}
    \zhhere_{AB} =  \gammexn_{AB}-\overline{\gammexn}_{AB}
     \,. 
  \end{equation}
Hence  
\begin{align}
H[\hyp,X,\backg](g)  = \  & 
- \frac{n}{16 \pi}
 \int_{\partial \hyp} 
      \zhhere_{AB}   
 ( \zD^{A}t )\zX^B  
    \sqrt{|\det \zgamma|} d^{n-1} x 
     \label{2IV26.3}
\\      = \  & 
\frac{n}{16 \pi}
 \big( 
  \ourQ[\hyp\cap\scri,X](g)
  -
  \ourQ[\hyp\cap\scri,X](\backg)
 \big)
  \label{2VI26.1}
\\     
  = \  & 
 \big( 
  \theirQ[\hyp\cap\scri,X](g)
  -
  \theirQ[\hyp\cap\scri,X](\backg)
 \big)\label{26VIII25.1f}
 \,,
\end{align}
as claimed.
This also makes it clear in which sense  \eqref{holcharge} 
  or \eqref{skencharge} show up  in a Hamiltonian analysis.

%% file: Comments.tex
\section{Which vectors $X$ and metrics  $\backg$?}
 \label{s2IV26.1}

Some comments about the conditions set so far are in order.

In the Hamiltonian calculations of \cite{ChAIHP} 
it has been assumed that the vector field $X$ is a Killing vector of $\backg$.  This condition was useful to analyse the convergence of the integrals. Moreover, solutions such as Minkowski spacetime or Anti-de Sitter spacetime provide natural asymptotic backgrounds for many metrics of interest, and therefore the condition seemed to be natural in the problem at hand at the time.

So let us consider the question of  convergence of the integrals appearing in this work. It follows from the calculations above that \emph{the equalities 
\eqref{2IV26.3}-\eqref{26VIII25.1f}  hold for any pair of vacuum metrics $g$ and $\backg$ which share the same conformal metric at $\scri$, and for all vectors $X$ extending smoothly to the conformal boundary of the spacetime,
with  the right-hand side of \eqref{26VIII25.1f} being  finite for such vectors}. 
This   takes care of the question of the convergence of  the integral defining $H$, and shows that neither $\backg$-Killing vectors in spacetime, nor conformal Killing vectors of $\scri$, are needed for \eqref{26VIII25.1f}.
 
However, a definition is only useful if it carries useful information. This is the
case when $X$ extends to  a conformal Killing vector of the conformal
boundary and when one of the tensors $\zht_{AB}$ or $\theirT_{AB}$ is divergence-free and trace-free, for then the associated charge integrals   are independent of the choice of a section of $\scri$. 
In such a case  a metric $\backg$ for which $X$ is a Killing vector which is asymptotically timelike, whenever  there is one, provides a natural reference value of energy.  As an example, one can prove positivity of $H$ for suitable classes of asymptotically Birmingham-Kottler metrics~\cite{Wang,ChHerzlich}, and for asymptotically Horowitz-Myers metrics~\cite{BrendleHung}, and for small perturbations of static $\backg$'s which themselves are small perturbations of the Anti-de Sitter metric~\cite{ChWuttePRL}. 
Whether or not the charges $\ourQ$, or $\theirQ$, or $H$, provide useful information in more general situations remains to be seen.